\documentclass[%
aps,preprint,prc,amsmath,amssymb,nofootinbib,superscriptaddress,11pt%
]{revtex4}

\usepackage[utf8]{inputenc}
\usepackage[usenames]{color}
\usepackage[dvipsnames]{xcolor}
\usepackage{graphicx}
\usepackage{amsmath}
\usepackage{amsfonts}
\usepackage{amssymb}
\usepackage{mathrsfs}
\usepackage{bm}
\usepackage{verbatim}
\usepackage{cancel}
\usepackage{comment}
\usepackage[normalem]{ulem}
\usepackage{CJK}
\usepackage{float}
\usepackage{textcomp}
\usepackage{booktabs}

\usepackage{hyperref}
\hypersetup{
  colorlinks=true,        
  linkcolor=ForestGreen,    
}

\usepackage{multirow,makecell}

\newcommand{\mlo}{M_{\text{lo}}}
\newcommand{\mhi}{M_{\text{hi}}}

\newcommand{\chn}[3]{{{}^{#1}\!{#2}_{#3}}}
\newcommand{\cs}[2]{\chn{#1}{S}{#2}}

\newcommand{\NNLO}{N$^2$LO}

\graphicspath{{./figs/}}

\begin{document}

 \begin{CJK*}{UTF8}{gbsn}

\title{Effective field theory with resonant $P$-wave interaction}

 \author{Qingfeng Li (李青峰)}
\affiliation{College of Physics, Sichuan University, Chengdu, Sichuan 610065, China}

\author{Songlin Lyu (吕松林)}
\email{songlin@scu.edu.cn}
\affiliation{College of Physics, Sichuan University, Chengdu, Sichuan 610065, China}

\author{Chen Ji (计晨)}
\email{jichen@ccnu.edu.cn}
\affiliation{Key Laboratory of Quark and Lepton Physics, Institute of Particle Physics,
Central China Normal University, Wuhan 430079, China}
\affiliation{Southern Center for Nuclear-Science Theory (SCNT), Institute of Modern Physics, Chinese Academy of Sciences, Huizhou 516000, Guangdong Province, China}

 \author{Bingwei Long (龙炳蔚)}
\affiliation{College of Physics, Sichuan University, Chengdu, Sichuan 610065, China}
\affiliation{Southern Center for Nuclear-Science Theory (SCNT), Institute of Modern Physics, Chinese Academy of Sciences, Huizhou 516000, Guangdong Province, China}

\date{June 25, 2023}

\begin{abstract}
A new effective field theory is developed to describe shallow $P$-wave resonances using nonlocal, momentum-dependent two-body potentials. This approach is expected to facilitate many-body calculations and is demonstrated to converge and to be renormalizable in perturbative calculations at subleading orders. The theory is applied to the neutron-$\alpha$ system, with good agreement found between its predictions and a phase-shift analysis of neutron-$\alpha$ elastic scattering. In the three-body system consisting of two neutrons and an $\alpha$ particle, the nonlocal potential in this framework is found to recover the same qualitative features as previously shown with energy-dependent formulations.
\end{abstract}

\maketitle

\section{Introduction\label{sec:intro}}

In the usual framework of nonrelativistic effective field theory (EFT) where  $t$-channel exchanged particles are integrated out, the momentum-independent contact operators are leading-order (LO) two-body interactions. If treated nonperturbatively, this formulation can be used to describe $S$-wave dynamics featuring a large scattering length. A series of momentum-polynomial terms will follow in perturbative calculations to produce corrections to the $S$ matrix in the form of the effective range. If the effective range is rather large~\cite{Beane:1997pk, Gegelia:1998gn, Bedaque:2003wa, Habashi:2020qgw, Habashi:2020ofb, Peng:2021pvo, Beane:2021dab, vanKolck:2022lqz} or a shallow $P$-wave resonance is present~\cite{Bertulani:2002sz, Bedaque:2003wa, Epelbaum:2021sns}, a single-parameter LO interaction becomes inadequate. An auxiliary dimeron field ~\cite{Bertulani:2002sz, Bedaque:2003wa, Long:2013cya} is often introduced to construct energy-dependent interactions so that a second fine-tuned parameter can be accounted for at LO. Other energy-dependent formulations without dimeron fields exist too~\cite{Epelbaum:2021sns}. However, energy-dependent potentials are not straightforward to implement in many-body calculations. For instance, when studying ${}^6$He with an energy-dependent neutron-alpha ($n \alpha$) potential~\cite{Rotureau:2012yu, Ji:2014wta, Ryberg:2017tpv}, one has to modify the standard orthogonality and closure conditions in order to retain hermiticity~\cite{Gobel:2019jba}. It is the goal of this paper to develop momentum-dependent EFT interactions for shallow $P$-wave resonances.

For the case where the $S$-wave scattering length and effective range are both large, an unconventional EFT framework was proposed in Refs.~\cite{Peng:2021pvo, Beane:2021dab, vanKolck:2022lqz} that can fine tune the scattering length and effective range simultaneously. At the center of the framework is the following nonlocal $S$-wave potential:
\begin{equation}
  V_0 (p^\prime, p) = -\frac{2\pi}{\mu} \frac{\lambda}{\sqrt{p^{\prime 2} + 2\mu \Delta}\sqrt{p^2 + 2\mu \Delta}}\, ,
  \label{eqn:pdep1s0}
\end{equation}
where $p$ ($p^\prime$) is the magnitude of the incoming (outgoing) center-of-mass (c.m.) momentum and $\mu$ the reduced mass. It is unusual to organize EFTs around such a nonlocal potential, but it has been shown for $NN$ scattering that subleading interactions can be systematically added while satisfying renormalization-group (RG) invariance~\cite{Beane:2021dab}. Not only does it recover the effective range expansion (ERE)~\cite{Beane:2021dab}, the potential can be used to accompany pion-exchanges in such a way chiral symmetry of the chiral Lagrangian is respected~\cite{Peng:2021pvo}. Applications of this potential to nuclear-structure calculations can be found in Refs.~\cite{SanchezSanchez:2020kbx,Yang:2020pgi}. 

We discuss in this paper how to extend the potential \eqref{eqn:pdep1s0} to describe a shallow $P$-wave resonance. The immediate application is the $n \alpha$ system. EFT construction for $n \alpha$ interactions is useful for describing model-independently the helium isotope chain and for shedding light on studies of other isotopes near the neutron drip line.

The LO $P$-wave potential is discussed in Sec.~\ref{sec:pwaveSEP}, with attention given to its differences in comparison with the energy-dependent counterpart. This is followed by construction of higher-order $P$-wave amplitudes in Sec.~\ref{sec:highorder}. Section~\ref{sec:3body} shows that the $n n \alpha$ three-body force is still at LO for renormalization purpose and finally, we summarize the findings in Sec.~\ref{sec:summary}.

\section{Nonlocal $P$-wave potential\label{sec:pwaveSEP}}

The obvious extension of the $S$-wave potential \eqref{eqn:pdep1s0} to $P$ waves is given by
\begin{equation}
  V^{(0)} (p^\prime, p) = -\frac{2\pi}{\mu} \frac{\lambda p' p}{\sqrt{p^{\prime 2} + 2\mu\Delta}\sqrt{p^2 + 2\mu\Delta}}\, ,
  \label{eqn:pwaveLO}
\end{equation}
where $\Delta > 0$. We would like to cast this potential into a Lagrangian. Since the $n \alpha$ system is the main application in the paper, it is chosen for presentation. With the two-component spinor $n$ (scalar $\alpha$) representing the neutron ($\alpha$) field and a four-component spinor $\Psi$ the auxiliary field for the $n\alpha$ $P_{3/2}$ resonance, we have
\begin{equation}
\begin{split}
    \mathcal{L}(x) &= 
    \frac{1}{2} n^\dagger \left(i\partial_0 + \frac{\nabla^2}{2m_n} \right) n + \alpha^\dagger \left(i\partial_0 + \frac{\nabla^2}{2m_\alpha} \right) \alpha + \frac{\mu}{\lambda} \Psi^\dagger \Psi \\
    & + \left[\Psi_a^\dagger(x) \int d^3 r \mathcal{F}(r)\,  n^T\left(x_0, \vec{x} + \frac{4}{5}\vec{r} \right) \vec{T}_a \cdot \hat{r}\,  \alpha\left(x_0, \vec{x} - \frac{\vec{r}}{5}\right) + \mathrm{H.c.} \right] + \cdots \, , \label{eqn:lag0}
\end{split}
\end{equation}
where $T_a^{i \,s} \equiv \langle 3/2, a | 1, i; 1/2, s\rangle$ are Clebsch-Gordan coefficients, $i$ is the Cartesian index for the three-dimensional unit vector, and $s$ the index for two-component spinors. Unlike in previous works employing the auxiliary field, $\Psi$ does not have a kinetic term, not even at higher orders, and thus does not propagate in time. Its coupling to $n \alpha$ is not point-like, instead underpinned by the spatial form-factor function 
\begin{equation}
    \mathcal{F}(r) = \frac{d}{dr} \int \frac{d^3 p}{(2\pi)^3} \frac{e^{-i\vec{p}\cdot\vec{r}}}{\sqrt{\vec{p}\,^2 + 2\mu \Delta}} \, .
\end{equation}
$n \alpha$ potentials are defined as Feynman diagrams irreducible by cutting $n \alpha$ intermediate states. The nonlocal $n\alpha \Psi$ vertex, combined with the instantaneous $\Psi$ propagator, eventually gives rise to the LO potential~\eqref{eqn:pwaveLO}.

One iterates this LO potential nonperturbatively by resumming all the $n \alpha$ bubbles, equivalent to solving the following partial-wave Lippmann-Schwinger equation (LSE):
\begin{equation}
   T_l(p^\prime, p; k) =  V_l(p^\prime, p) + \frac{\mu}{\pi^2} \int^\Lambda \mathrm{d} q \, q^2\, V_{l}(p^\prime, q) \frac{T_l(q, p; k)}{k^2 - q^2 + i0} \, ,
   \label{eqn:LSE}
\end{equation}
where $k^2/2\mu$ is the c.m. energy. Here $\Lambda$ only serves the purpose of indicating that the integral is regularized by a momentum cutoff. The regulator function does not have to be a sharp cutoff, but it is assumed to be at least separable. The separable form of $V^{(0)}(p', p)$ facilitates the straightforward solution
\begin{equation}
    T^{(0)}(p', p; k) = -\frac{2\pi}{\mu} \frac{p'}{\sqrt{p^{\prime 2} + 2\mu\Delta}}\, \tau(k)\, \frac{p}{\sqrt{p^2 + 2\mu\Delta}}  \, ,    
\end{equation}
where
\begin{equation}
    \begin{split}
    \tau(k) &= \left[\lambda^{-1} + \frac{2}{\pi} \int^\Lambda dq \frac{q^4}{(q^2 + \gamma^2)(k^2 - q^2 + i0)}  \right]^{-1} \\
    &= \frac{k^2 + \gamma^2}{\gamma^2 (\gamma + \lambda_R^{-1}) +  \lambda_R^{-1} k^2 - i k^3}     
    \end{split}\label{eqn:taudef}
\end{equation}
with renormalized $\lambda_R$ and $\gamma$ defined by
\begin{align}
    \lambda_R^{-1} &\equiv \lambda^{-1} - \frac{2}{\pi} \int^\Lambda dq \, , \\
    \gamma &\equiv \sqrt{2\mu \Delta} > 0\, .
\end{align}
It could be useful to make it clear that $\tau(k)$ has generally two poles by rewriting it as
\begin{equation}
    \tau(k) = \frac{ik - \gamma }{k^2 + i\eta k - \eta \gamma} \, , \label{eqn:tauk}
\end{equation}
thus presenting the full off-shell $T^{(0)}(p', p; k)$ instead as
\begin{equation}
    T^{(0)}(p', p; k) = -\frac{2\pi}{\mu} \frac{p'}{\sqrt{{p'}^2 + \gamma^2}} \frac{ik - \gamma }{k^2 + i\eta k - \eta \gamma} \frac{p}{\sqrt{{p}^2 + \gamma^2}} \, .\label{eqn:offT_2}
\end{equation}
where $\eta \equiv \gamma + \lambda_R^{-1}$. Although the off-energy-shell amplitude has branch cuts in the complex plane of $p$ or $p'$, the on-shell LO amplitude does not and it has the wanted ERE of the $T$ matrix:
\begin{equation}
\begin{split}
    T^{(0)}(k, k; k) &= -\frac{2\pi}{\mu} \frac{k^2}{k^2+\gamma^2} \tau(k) \\
    &= -\frac{2\pi}{\mu} \frac{k^2}{\gamma^2 \eta + \lambda_R^{-1} k^2 - i k^3} 
    \, ,
    \label{eqn:LOT1}    
\end{split}
\end{equation}
where the ERE parameters are identified as
\begin{align}
    a_1^{-1} &= - \gamma^2 \eta \, , \\
    \frac{r_1}{2} &= \lambda_R^{-1} \, .
\end{align}

For this potential to support an EFT, it is crucial to show that systematic corrections can be added on top of it. We will investigate in Sec.~\ref{sec:highorder} what subleading potentials lead to higher ERE terms. Before that, we discuss what kind of two-body physics can be described by the LO amplitude if $\lambda$ and $\Delta$ are varied.

$p = i\gamma$ and $p' = i\gamma$ are a branch point of off-shell $T^{(0)}(p', p; k)$~\eqref{eqn:offT_2}, and they combine to become a pole when $p = p' = k$ in Eq.~\eqref{eqn:LOT1}. [The other branch point $p (p') = -i \gamma$ cancels the zero of $\tau(k)$ when $p = p' = k$.] 
Interestingly, the pole at $k = i\gamma$ does not correspond to a physical bound state. In fact, the bound states, if any, must be associated with the poles of $\tau(k)$. To see this, we begin by recalling that the Schr\"odinger equation for the bound states is equivalent to the homogeneous LSE:
\begin{equation}
    (H_0 + V) | \Psi \rangle = E | \Psi \rangle \Longrightarrow | \Psi \rangle = \frac{1}{E - H_0} V | \Psi \rangle \, ,
\end{equation}
where $H_0$ is the kinetic energy, $V$ the $P$-wave potential~\eqref{eqn:pwaveLO}, $E$ the energy eigenvalue and $| \Psi \rangle$ the eigenstate. In momentum space, this abstract equation becomes a homogeneous variant of Eq.~\eqref{eqn:LSE}, provided that one redefines $\Psi(q; E) \equiv T(q; E)/(2\mu E - q^2)$:
\begin{equation}
    \Psi(p; E) = - \frac{2}{\pi} \frac{1}{2\mu E - p^2} \frac{\lambda p}{\sqrt{{p}^2 + \gamma^2}}  \int^\Lambda \mathrm{d} q \, q^2\, \frac{q}{\sqrt{q^2 + \gamma^2}} \Psi(q; E)  \, .
   \label{eqn:LSE_Homo}
\end{equation}
Because the integral on the right-hand side does not depend on $p$, the wave function must have the form
\begin{equation}
    \Psi(p; E) \propto \frac{1}{2\mu E - p^2} \frac{p}{\sqrt{{p}^2 + \gamma^2}} \, ,
\end{equation}
and the eigenvalue $E$ satisfies
\begin{equation}
   \lambda\left[ \lambda^{-1} + \frac{2}{\pi} \int^\Lambda dq \frac{q^4}{(q^2 + \gamma^2)(2\mu E - q^2)} \right] = 0\, .
\end{equation}
Comparing the above equation with Eq.~\eqref{eqn:taudef}, we see that an energy eigenvalue is precisely the location of a pole of $\tau(\sqrt{2\mu E})$. This also agrees with what is expected from the Lehmann-Symanzik-Zimmermann reduction formula: the full off-shell $T^{(0)}(p', p; k)$  
factorizes as $k \to \sqrt{2\mu E}$ while $p'$ and $p$ stay fixed,
\begin{equation}
    T^{(0)}(p', p; k) \to \frac{p'}{\sqrt{{p'}^2 + \gamma^2}} \frac{R(E)}{k^2/2\mu - E} \frac{p}{\sqrt{{p}^2 + \gamma^2}} + \, \text{finite terms} \, .\label{eqn:TFactorize}
\end{equation}
Comparing Eq.~\eqref{eqn:offT_2} with Eq.\eqref{eqn:TFactorize}, we find that the $k$-plane pole term $R(E)/(k^2/2\mu - E)$ can only be attributed to $\tau(k)$.

This loss of correspondence between a positive imaginary $k$-plane pole of the $S$ matrix and a bound state is unusual, but it could be advantageous. In the energy-dependent potential implemented by the dynamically propagated dimeron field~\cite{Braaten:2011vf, Nishida:2011np}, the third pole corresponds to a negative-norm state, which eventually causes to violate the Wigner bound~\cite{Hammer:2009zh, Hammer:2010fw}. 

With that, we conclude that $P$-wave threshold dynamics is primarily decided by the pair of poles of $\tau(k)$, denoted by $k_\pm$. We now categorize the poles according to the values taken by $\lambda$ and $\gamma$. Because we are interested in only attractive forces, the bare parameter $\lambda$ is always positive, and $\gamma$ is positive too by construction. Depending on the value of $\eta = \gamma + \lambda_R^{-1}$, the poles $k_\pm$ can be on the positive imaginary axis (bound state), on the negative imaginary axis (virtual), or in the lower half-plane (resonance), as tabulated in Table~\ref{tab:pole_T0}. Except for in the row labeled by ``bound / virtual'' where one pole is virtual state and the other is bound state, the pair of poles are either both resonance or both virtual.

\begin{table}
\renewcommand{\arraystretch}{1.5}
  \centering
    \begin{tabular}{c c c}
    \hline\hline
     Type & $\eta$ & Pole position $(k_\pm)$ \\
    \hline
    bound / virtual &  
     $\eta < 0$ & \quad $\frac{i}{2} \left[ |\eta| \pm \sqrt{|\eta| (4\gamma + |\eta|)} \right]$ \\
    \hline
    \multirow{2}{*}{resonance}  & $0 < \eta$ & $ \quad \frac{1}{2} \left[ \pm \sqrt{\eta \left(3\gamma + |\lambda_R^{-1}| \right)} - i \eta \right]$  \\
    & $0 < \eta < 4\gamma$ & $ \quad \frac{1}{2} \left(\pm \sqrt{4\eta \gamma - \eta^2} - i\eta  \right)$  \\
    \hline
    virtual &  $4\gamma < \eta$ & \quad $ -\frac{i}{2} \left(\eta \pm \sqrt{\eta^2 - 4\eta \gamma} \right)$ \\
    \hline\hline
\end{tabular}
    \caption{Categories of pole positions of $\tau(k)$ Eq.~\eqref{eqn:tauk} according to the value of $\eta$.}
    \label{tab:pole_T0}
\end{table}

One would usually assume $\gamma$ and $\lambda_R^{-1}$ to be independent low-energy parameters, and refer to them by a generic low-energy scale $\mlo$. Therefore, the position of $k_+$ and $k_-$ is comparable with $\mlo$. But a less boring scenario is realized by nature: In the case of $n\alpha$ $P$-wave resonance~\cite{Bertulani:2002sz, Bedaque:2003wa, vanKolck:2004te}, the real part of the resonance position is over five times as large as the imaginary part. Or equivalently, $|k_+ - k_-|/|k_+ + k_-| \simeq 5.6$, with $k_\pm = (\pm 34.5 - i 6.2)$ MeV obtained from the $P_{3/2}$ ERE parameters of $n\alpha$ scattering: $a_1 = -62.951\, \text{fm}^3$ and $r_1 = -0.8819\, \text{fm}^{-1}$~\cite{Arndt:1973ssf}. As can be seen from Table~\ref{tab:pole_T0}, the smallness of their imaginary part comes to life because of near cancellation in $\eta = \gamma + \lambda_R^{-1}$. Reference~\cite{Bedaque:2003wa} chose accordingly the following scaling for low-energy constant (LECs):
\begin{align}
    \lambda_R^{-1} &= -87.0 \, \text{MeV} \sim \mhi  \, , \\
    \gamma &= 99.4 \, \text{MeV} \sim \mhi \, , \\
    \eta &= 12.4\, \text{MeV} \sim \frac{\mlo^2}{\mhi} \, .
\end{align}
This arrangement actually makes it ``easier'' for the underlying theory to form such a shallow resonance in the sense only $\eta$ needs to be fine-tuned. It also sets the breakdown scale $\mhi \simeq 90$ MeV. Although this special scaling is interesting, we continue to count $\gamma$ and $\lambda_R^{-1}$ as low-energy scales for the rest of the paper. When $\gamma$ or $\lambda_R^{-1}$ becomes large, one needs to expand the results in $\gamma^{-1}$ or $\lambda_R$ and drop higher-order terms.

\section{Higher orders\label{sec:highorder}}

We wish to show that this formulation can be systematically improved. In analogy to adding derivative couplings for subleading corrections in many effective field theories, we add derivatives to the form-factor function in Eq.~\eqref{eqn:lag0} and construct higher-dimension $n\alpha \Psi$ transition vertex
\begin{equation}
\begin{split}
    \mathcal{L}^{(1)}(x) = \frac{g_2}{2 \lambda} \left[\Psi_a^\dagger(x) \int d^3 r\,  n^T\left(x_0, \vec{x} + \frac{4\vec{r}}{5}\right) \vec{T}_a \cdot \hat{r} \frac{d^2 \mathcal{F}(r)}{dr^2} \alpha\left(x_0, \vec{x} - \frac{\vec{r}}{5}\right) + \mathrm{H.c.} \right] \, , \label{eqn:lag1}
\end{split}    
\end{equation}
which translates to the following next-to-leading order (NLO) $n \alpha$ potential:
\begin{equation}
    V_{g_2}^{(1)}(p', p) = \frac{2\pi}{\mu}\, \frac{g_2}{2} \,  \frac{p' p\, (p'^2 + p^2)}{\sqrt{p'^2 + \gamma^2}\sqrt{p^2 + \gamma^2}}  \, .
    \label{eqn:Vbeta1}
\end{equation}
In addition, we need to account for corrections to $\lambda$ and $\Delta$ at subleading orders, which is implemented by formally expanding the bare parameters,
\begin{eqnarray}
    \lambda & = & \lambda^{(0)} + \lambda^{(1)} + \cdots \, , \label{eqn:DeltaExp} \\
    \Delta & = & \Delta^{(0)} + \Delta^{(1)} + \cdots \, , \label{eqn:LambdaExp}
\end{eqnarray}
and using this expansion in Eq.~\eqref{eqn:pwaveLO}:
\begin{equation}
\begin{split}
    V^{(1)}_\lambda(p', p) &= -\frac{2\pi}{\mu} \frac{p^\prime p}{\sqrt{p^{\prime 2} + 2\mu\Delta^{(0)}} \sqrt{p^2 + 2\mu\Delta^{(0)}}}\\
    &\qquad \times \left\{ \lambda^{(1)} - \lambda^{(0)} \mu\Delta^{(1)} \left(\frac{1}{p^{\prime 2} + 2\mu\Delta^{(0)}} + \frac{1}{p^2 + 2\mu\Delta^{(0)}} \right) \right\} \, .
\end{split}
\end{equation}
The sum of $V_{g_2}^{(1)}$ and $V_\lambda^{(1)}$ resembles the NLO $S$-wave potential in Ref.~\cite{Beane:2021dab} except for the $P$-wave momenta prefactor $p' p$.

The justification for using these NLO potentials comes ultimately from the generalized shape parameter they produce\,---\,$P_1 k^4$. Treating $V_{g_2}^{(1)}$ as a perturbation, one obtains its correction to the $T$ matrix:
\begin{equation}
    \begin{split}
    T^{(1)}_{g_2} (k;\Lambda) = & \frac{\mu}{2\pi}\frac{\left[ T^{(0)}(k) \right]^2}{k^2} \frac{g_2}{\lambda} \left[\gamma^2 L_{\Lambda 3} + \left(L_{\Lambda 3} + \gamma^2 \lambda^{-1}\right)k^2 + \lambda^{-1} k^4 \right] \, ,
    \end{split} \label{eqn:OneInsG2}
\end{equation}
where
\begin{equation}
     L_{\Lambda 3} \equiv \gamma^3 + \frac{2}{\pi}\int^\Lambda dq (q^2 - \gamma^2)  \, .
\end{equation}
Here, we have dropped the superscript ${}^{(0)}$ of the LO parameters $\lambda$ and $\gamma$ to avoid cluttered symbols. Insertion of $V_\lambda^{(1)}$ into the LO amplitude provides necessary counterterms to absorb divergences in $T^{(1)}_{g_2}$. The renormalized NLO amplitude is then given by
\begin{equation}
    \begin{split}
    T^{(1)} (k) = &-\frac{\mu}{2\pi}\frac{\left[ T^{(0)}(k) \right]^2}{k^2} \left[\frac{\lambda^{(1)}_R}{\lambda_R^2}\gamma^2 - 2\mu\Delta^{(1)}\left(\lambda_R^{-1} + \frac{3}{2}\gamma \right) -\frac{g_{2R}}{\lambda_R}\gamma^5 \right.  \\
    &\left. + \left(\frac{\lambda^{(1)}_R}{\lambda_R^2} - \frac{g_{2R}}{\lambda_R}\gamma^2 \left(\lambda_R^{-1} + \gamma \right) \right)k^2  - \frac{g_{2R}}{\lambda_R^2} k^4 \right]\, ,
    \end{split}\label{eqn:TNLO}
\end{equation}
where the renormalized parameters are defined as
\begin{align}
    \frac{g_{2R}}{\lambda^2_R} &= \frac{g_{2}(\Lambda)}{\lambda^2(\Lambda)}\, , \\
    \quad 
    \frac{\lambda^{(1)}_R}{\lambda_R^2} - \frac{g_{2R}}{\lambda_R}\gamma^3 &= \frac{\lambda^{(1)}(\Lambda)}{\lambda^2(\Lambda)} - \frac{g_{2}(\Lambda)}{\lambda(\Lambda)}
    \left(\gamma^3 + \frac{2}{\pi} \int^\Lambda dq q^2 \right)\,.    
\end{align}
Again, we have dropped the superscript ${}^{(0)}$ of $\lambda_R$. The NLO corrections to ERE parameters are identified as follows:
\begin{align}
    a_1^{(1)} &=  \left[\frac{g_{2R}}{\lambda_R}\gamma^5 + 2\mu\Delta^{(1)}(\lambda_R^{-1} + \frac{3}{2}\gamma) - \frac{\lambda^{(1)}_R}{\lambda_R^2} \gamma^2 \right] \frac{1}{\gamma^4 (\gamma + \lambda_R^{-1})^{2}} \, , \\
     \frac{r_1^{(1)}}{2} &= -\frac{\lambda^{(1)}_R}{\lambda_R^2} + \frac{g_{2R}}{\lambda_R} \left(\lambda_R^{-1} + \gamma \right) \, , \\
    P_1 &=  \frac{g_{2R}}{\lambda_R^2} \, .
\end{align}
These expressions show that the NLO terms, either divergent or convergent with respect to $\Lambda$, can be rearranged into the form of the ERE series. It is important for $\lambda^{(1)}$ and $\Delta^{(1)}$ to appear in the corrections to $a_1$ and $r_1 k^2/2$ so that they can absorb divergences generated by the $g_2$ coupling.

At next-to-next-to-leading order ({\NNLO}), it becomes rather cumbersome to list all the analytic expressions that show how the LECs are renormalized and how they work together to construct a still higher-order ERE term $\propto k^6$. In addition, for the application to $n\alpha$ scattering the NLO already achieves excellent agreement with the phase shift analysis from Ref.~\cite{Arndt:1973ssf}, as we will see later. Therefore, instead of displaying the full set of {\NNLO} expressions we will be content to know that there are enough counterterms to absorb new divergences and that besides a $k^6$ ERE term there is no other analytic structure of the amplitude emerging at {\NNLO}, such as a pole or a branch cut.

There are three {\NNLO} contributions that are most divergent. The first stems from a new $n\alpha \Psi$ vertex with a fourth derivative:
\begin{equation}
\begin{split}
    \mathcal{L}^{(2)}(x) = \frac{g_4}{2 \lambda} \left[\Psi_a^\dagger(x) \int d^3 r\,  n^T\left(x_0, \vec{x} + \frac{4\vec{r}}{5}\right) \vec{T}_a \cdot \hat{r} \frac{d^4 \mathcal{F}(r)}{dr^4} \alpha\left(x_0, \vec{x} - \frac{\vec{r}}{5}\right) + \mathrm{H.c.} \right] \, , \label{eqn:lag2}
\end{split}    
\end{equation}
which gives rise to a {\NNLO} potential
\begin{equation}
    V_{g_4}^{(2)}(p', p) = \frac{2\pi}{\mu}\, \frac{g_4}{2} \,  \frac{p' p\, (p'^4 +  p^4)}{\sqrt{p'^2 + \gamma^2}\sqrt{p^2 + \gamma^2}} \, .
    \label{eqn:Vbeta2}
\end{equation}
Inserting this potential into the LO amplitude generates the following correction:
\begin{equation}
    \begin{split}
    T^{(2)}_{g_4} (k; \Lambda) = & \frac{\mu}{2\pi}\frac{\left[ T^{(0)}(k) \right]^2}{k^2} \frac{g_4}{\lambda} \left[ \gamma^2 L_{\Lambda 5} + k^2 \frac{2}{\pi}\int^\Lambda dq q^4 \right. \\
    & + \left. (L_{\Lambda 3} + \lambda^{-1}\gamma^2)k^4 + \lambda^{-1}k^6 \right]\, ,
    \end{split}\label{eqn:TNNLO2-g4}
\end{equation}
where 
\begin{equation}
    L_{\Lambda 5} \equiv -\gamma^5 + \frac{2}{\pi}\int^\Lambda dq (q^4 - q^2\gamma^2 + \gamma^4)  \, .
\end{equation}

A second {\NNLO} potential has no free parameters, constructed by connecting two NLO $n\alpha\Psi$ vertexes with a $\Psi$ propagator:
\begin{equation}
    V_{g_2\Psi g_2}^{(2)}(p', p) = -\frac{2\pi}{\mu}\, \frac{g_2^2}{4 \lambda} \,  \frac{p' p\, (p'^2  p^2)}{\sqrt{p'^2 + \gamma^2}\sqrt{p^2 + \gamma^2}} \, .
    \label{eqn:Vbeta1sqr}
\end{equation}
(It might be possible to use the equation of motion to reduce this potential to the $g_4$ one~\cite{Beane:2000fi}, but it is rather a digression from our main line of investigation.) The corresponding contribution to the amplitude is given by
\begin{equation}
    \begin{split}
    T^{(2)}_{g_2\Psi g_2} (k;\Lambda) =& -\frac{\mu}{2\pi}\frac{\left[ T^{(0)}(k) \right]^2}{k^2} \frac{g_{2}^2}{4 \lambda} \left[ \gamma^2 L_{\Lambda 3}^2 \right. \\ 
    & \left. + (L_{\Lambda 3} + 2\gamma^2\lambda^{-1})L_{\Lambda 3}k^2
    + 2\lambda^{-1} L_{\Lambda 3} k^4 + \lambda^{-2}k^6 \right] \, .
    \end{split}\label{eqn:TNNLOG2G2}
\end{equation}
This {\NNLO} amplitude is not to be confused with two insertions of $V_{g_2}^{(1)}$ into the LO diagrams. In the latter, the neighboring $g_2$ vertexes are connected by a pair of $n \alpha$ propagators instead of a $\Psi$ propagator:
\begin{equation}
    \begin{split}
        T^{(2)}_{g_2} (k;\Lambda) = & \frac{\mu}{2\pi}\frac{\left[ T^{(0)}(k) \right]^2}{k^2} \left(\frac{g_2}{2\lambda} \right)^2 \Big[\gamma^2\left(\lambda L_{\Lambda 3}^2 - L_{\Lambda 5}\right)  \\
        & \qquad + \left(\lambda L_{\Lambda 3}^2 - L_{\Lambda 5} + 5\gamma^2 L_{\Lambda 3} \right)k^2  \\
        & \qquad + \left(5L_{\Lambda 3} + 4\lambda^{-1}\gamma^2 \right)k^4 + 4\lambda^{-1} k^6 \Big]  \\
        & + \left(\frac{\mu}{2\pi}\right)^2 \frac{\left[ T^{(0)}(k) \right]^3}{k^4} \left(\frac{g_2}{\lambda} \right)^2  \\
        & \qquad \times \Big[ \gamma^2 L_{\Lambda 3} + (L_{\Lambda 3} + \lambda^{-1}\gamma^2)k^2 + \lambda^{-1}k^4 \Big]^2 \, .
    \end{split} \label{eqn:TwoInsG2}
\end{equation}
We have split $T^{(2)}_{g_2}$ into two parts: one is proportional to $[T^{(0)}]^2$ and the other to $[T^{(0)}]^3$. The $[T^{(0)}]^3$ part contains no new physics compared with Eq.~\eqref{eqn:OneInsG2}, as it merely restores unitarity at the $\mathcal{O}(g_2^2)$ level. The $[T^{(0)}]^2$ part, however, indeed provides new corrections to ERE terms up to $k^6$. 

Yet another {\NNLO} potential is constructed by collecting second-order terms in expansions ~\eqref{eqn:DeltaExp} and \eqref{eqn:LambdaExp}, such as $\lambda^{(2)}$, $\Delta^{(2)}$, and $\Delta^{(1)}\lambda^{(1)}$. Two insertions of the NLO potential $V_\lambda^{(1)} + V_{g_2}^{(1)}$ also contribute. They are expected to generate corrections to $a_1$, $r_1 k^2/2$, and $P_1 k^4$, and these corrections, together with $g_4$, are able to absorb divergences exhibited as ERE coefficients up to $k^6$ in Eqs.~\eqref{eqn:TNNLO2-g4}, \eqref{eqn:TNNLOG2G2}, and \eqref{eqn:TwoInsG2}.

\begin{figure}
    \centering
    \includegraphics[width = 0.45\textwidth]{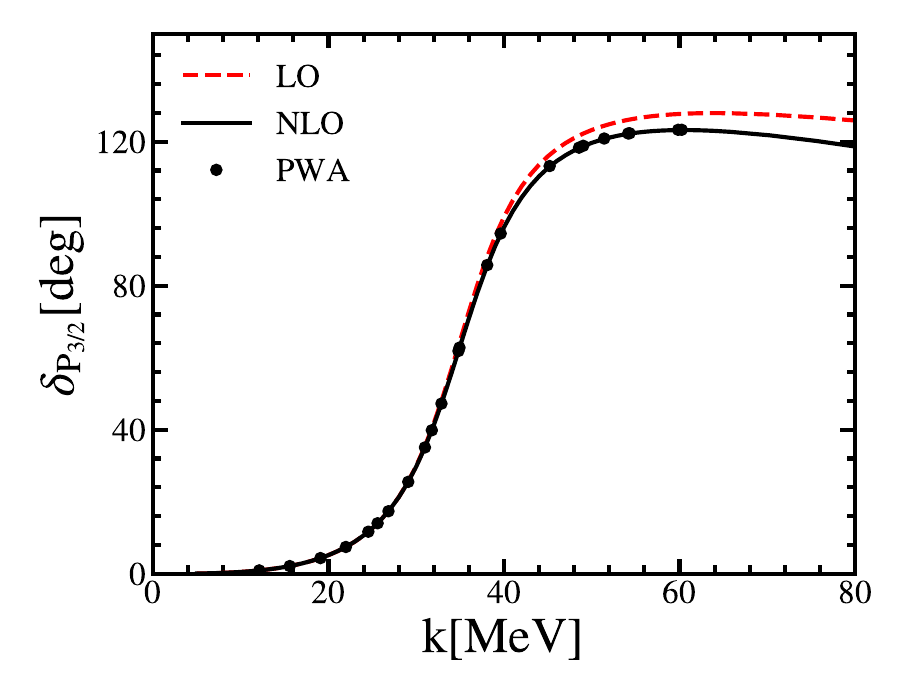}
    \caption{The $P_{3/2}$ phase shifts of $n\alpha$ scattering as a function of $k$. ``PWA'' refer to the phase shift analysis from Ref.~\cite{Arndt:1973ssf}.
    }
    \label{fig:phaseshift_3}
\end{figure}

We put these EFT potentials to the test by applying them to $n \alpha$ scattering. The LS equation is solved numerically, with the LO potential fully iterated while NLO and {\NNLO} potentials treated perturbatively. In the calculations, the potentials are regularized by the quartic Gaussian function:
\begin{equation}
    V(p', p) \to \exp\left(-\frac{p^4}{\Lambda^4}\right) V(p', p) \exp\left(-\frac{{p'}^4}{\Lambda^4}\right) \, .
\end{equation}
In Fig.~\ref{fig:phaseshift_3}, the $P_{3/2}$ phase shifts of $n\alpha$ scattering as a function of $k$ are compared with the phase shift analysis~\cite{Arndt:1973ssf}. The LECs are obtained by fitting to empirical values below and around the resonance peak $k \simeq 40$ MeV. Because we have demonstrated renormalization analytically up to NLO, therefore, only one value of $\Lambda$ is used: $\Lambda = 0.8$ GeV. Already at NLO the EFT achieves sufficient accuracy and leaves little room to improve.

\section{Three-body system of neutron-neutron-$\alpha$
\label{sec:3body}}

If the two-body system is the only application, developing an EFT to reproduce the well-known ERE seems an overkill. Nuclear EFTs often find their usefulness when more particles are present, e.g., a photon or other nucleons. Having built the EFT potentials for $n\alpha$, we apply its LO term in this section to the three-body system of $nn\alpha$~\cite{Rotureau:2012yu, Ji:2014wta, Ryberg:2017tpv}. 
This application is part of the research program of so-called halo/cluster EFTs (see Refs.~\cite{vankolck:2005ijmpe, Hammer:2017tjm,Ando:2020wtu, Hammer:2019poc} for reviews). 

The $nn\alpha$ system is interesting in its own right. In this three-body system, none of the two-body subsystems are bound but $nn\alpha$ together can form the stable isotope of $^6$He, an instance of so-called Borromean states. The $n\alpha$ interaction supports a narrow $P$-wave resonance but not a bound state, and the $nn$ pair is not known to be bound either, dictated instead by a shallow $S$-wave virtual state. However, the three-body system is bound by 0.969 MeV, known as the ground state of $^6$He. 

We follow mostly the computational framework of Refs.~\cite{Ji:2014wta, Hammer:2017tjm}, with the only difference being the form of $n\alpha$ interaction. Some of the technicalities will be reviewed briefly. In addition to Eq.~\eqref{eqn:lag0}, we need Lagrangian terms to describe $nn$ interaction and, especially, the short-range three-body force that couples a neutron spectator to the $n\alpha$ pair in the $P$ wave:
\begin{align}
    \mathcal{L}_{nn} &= -C_{\phi n} \left( \phi^\dagger n_\delta S^\delta n_{-\delta} + \mathrm{H.c.} \right) + \sigma \phi^\dagger \phi \,, \\
    \mathcal{L}_{nn\alpha} &= - h \left(G_0^{ab} T_a^{is} \Psi_b \overleftrightarrow{\partial_i} n_s \right)^\dagger \left( G_0^{cd} T_c^{jr} \Psi_d \overleftrightarrow{\partial_j} n_r\right) \, .
\end{align}
where $\phi$ is the $\cs{1}{0}$ $nn$ dibaryon field, and $S^\delta \equiv \langle 0, 0 | 1/2, \delta; 1/2, -\delta\rangle$
and $G_0^{ab} \equiv \langle 0, 0 | 3/2, a; 3/2, b\rangle$ are the Clebsch-Gordan coefficients and dummy indices are assumed to be summed up. With $m_n$ ($m_\Psi$) the mass of the neutron ($n\alpha$ resonance)~\cite{Hammer:2017tjm},
$\overleftrightarrow{\partial_i} \equiv [(m_n \overleftarrow{\nabla} - m_\Psi\overrightarrow{\nabla})/( m_n + m_\Psi)]_i $ is the Galilean invariant derivative. 

\begin{figure}
\centerline{\includegraphics[width=14cm,angle=0,clip=true]{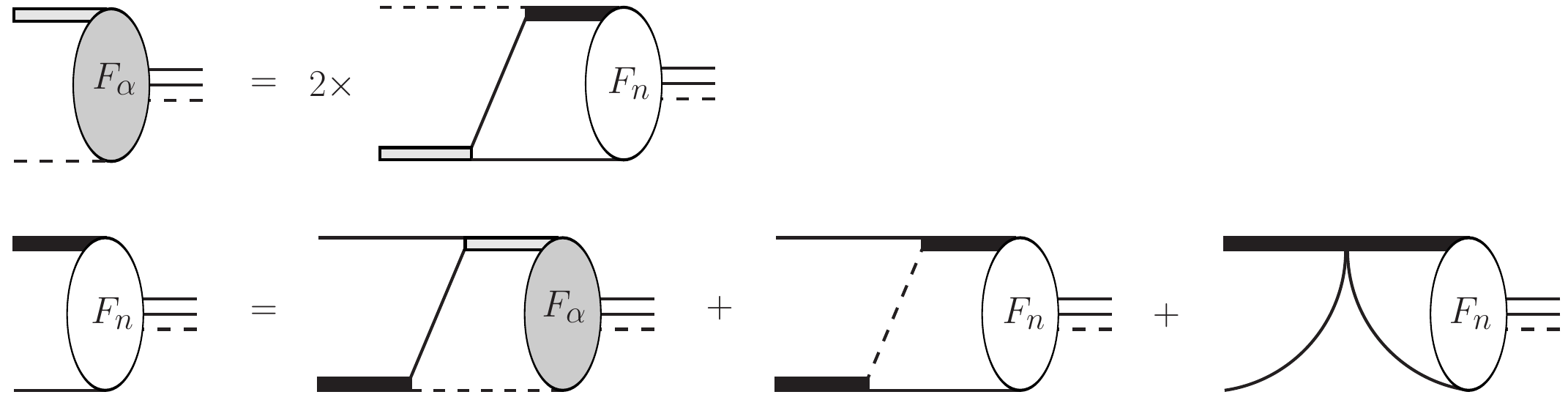}}
\caption{Diagrammatic representation for the coupled-channel $nn\alpha$ Faddeev equations.
The single dashed and solid lines refer, respectively, to the $\alpha$ and neutron field. The grey and dark thick lines represent respectively the $nn$ and $n\alpha$ dimeron fields, $\phi$ and $\Psi$. 
}
\label{fig:6he}
\end{figure}

The three-body equation for the $^6$He bound state can be described diagrammatically in Fig.~\ref{fig:6he}, where the two Faddeev components $F_\alpha$ and $F_n$ are defined according to which of $n$ and $\alpha$ is the spectator in Jacobi coordinates~\cite{Ji:2014wta,Hammer:2017tjm}. The spin-parity $0^+$ state of the entire three-body system is considered, while at LO the $n n$ pair interacts in $\cs{1}{0}$ and the $n\alpha$ pair in the $^2P_{3/2}$ resonant channel. $F_\alpha$ and $F_n$ are defined such that the relative Jacobian angular momentum between the spectator particle and the pair is projected onto the $S$ wave between $\alpha$ and the $nn$ pair, and onto the $P$ wave between $n$ and the $n\alpha$ pair. 

We use the second diagram of the second line in Fig.~\ref{fig:6he} to explain the kinematics. As a rule, the spectator external line is always chosen to be on-shell. Therefore, the neutron external line carries on-shell four-momentum $(q^2/2m_n, \vec{q}\,)$, and the $\Psi$ external line for the $n\alpha$ pair has off-shell four-momentum $(-B_3 - q^2/2m_n, -\vec{q}\,)$ so that the three-body total energy is $-B_3$. After the $n\alpha$ pair exchanging a neutron with the spectator neutron, the remaining $\alpha$ carries a final four-momentum of $(q_0', \vec{q}\,')$, while the emerging $nn$ pair has $(-B_3 - q_0', -\vec{q}\,')$. In addition, the $n\alpha$ pair carries a relative momentum between its constituents:
\begin{eqnarray}
\vec{\pi}_1 = \frac{4}{5}\vec{q}+\vec{q}\,' \, .
\end{eqnarray}
We integrate out $q_0'$ by picking the pole of the $\alpha$ propagator, which is consistent with the aforementioned rule of making the external spectator on-shell. The resulting integrals can be further decomposed by expanding with the Legendre polynomials in $\hat{q}\cdot\hat{q}'$. 

There are several key elements in assembling the coupled-channel integral equations, including the kernel functions and the dressed $nn$ and $n\alpha$ dimeron propagators. The kernel functions are the propagators of the exchanged particles multiplied by appropriate vertices. The one corresponding to the exchanged neutron combined with the $\phi n n$ and $n \alpha \Psi$ vertices, shown in the second diagram of the second line in Fig.~\ref{fig:6he}, is given by
\begin{eqnarray}
X_{n\alpha}(q,q';B_3) 
= \sqrt{2}\left[\frac{4}{5} q\, G_{n\alpha,0} (q,q';B_3) + q'\, G_{n\alpha,1} (q,q';B_3)\right]\, , \label{eq:faddeev-Xnalpha}
\end{eqnarray}
where the angle-averaged Green's function $G_{n\alpha, l}$ is projected onto the orbital quantum number $l$ by the Legendre polynomial $P_l$:
\begin{equation}
\begin{split}
G_{n\alpha,l} (q,q';B_3) = \frac{m_n}{2} \int_{-1}^{1} d (\hat{q}\cdot\hat{q}') \frac{P_l(\hat{q}\cdot\hat{q}')}{\sqrt{\pi_1^2+\gamma^2} \left(m_n B_3 + \frac{5}{8}\pi_1^2 + \frac{3}{5} q^2\right)} \, .
\end{split}
\end{equation}
The kernel function for the second diagram of the first line in Fig.~\ref{fig:6he} is similar to $X_{n\alpha}(q, q'; B_3)$, except for a simple adjustment of switching $q$ and $q'$. The kernel function for the exchanged $\alpha$ in the third diagram of the second line is given by
\begin{eqnarray}
X_{nn}(q,q';B_3) 
&=& \frac{27}{25} qq'\,G_{nn,0}(q,q';B_3) \nonumber \\
& &\; +\frac{2}{5} (q^2+q'^2)\,G_{nn,1}(q,q';B_3)
+ q q'\,G_{nn,2}(q,q';B_3),
\label{eq:faddeev-Xnn}
\end{eqnarray}
where 
\begin{eqnarray}
G_{nn,l} (q,q';B_3) = \frac{m_n}{2} \int_{-1}^{1} d (\hat{q}\cdot\hat{q}') \frac{P_l(\hat{q}\cdot\hat{q}')}{\sqrt{\pi_2^2+\gamma^2} \left(m_n B_3 + \frac{5}{8}\pi_2^2 + \frac{3}{5} q^2\right) \sqrt{\pi_3^2+\gamma^2}}\, ,
\label{eq:faddeev-Gnn}
\end{eqnarray}
with
\begin{eqnarray}
\vec{\pi}_2 &=&\frac{1}{5}\vec{q} + \vec{q}\,' \, ,
\nonumber\\
\vec{\pi}_3 &=& \vec{q} + \frac{1}{5}\vec{q}\,' \, .
\end{eqnarray}
For more detailed explanation concerning, for instance, the angular-momentum recoupling coefficients, we refer to Ref.~\cite{Ji:2014wta}.

The $nn$ interaction is subsumed in the dressed $\cs{1}{0}$ dibaryon propagator, depicted by the grey lines, that depends only on the $\cs{1}{0}$ scattering length $a_0 = -18.6$ fm at LO:
\begin{eqnarray}
D_{nn}(\kappa) = -\frac{1}{2\pi^2 m_n} \frac{1}{\kappa -a_0^{-1}}\, ,
\end{eqnarray}
where $\kappa$ is related to the dibaryon four-momentum $(p_0, \vec{p}\,)$ through $\kappa \equiv \sqrt{-m_n p_0 + \vec{p}\,^2 -i0}$. The $\cs{1}{0}$ $nn$ scattering amplitude can be written with $D_{nn}$ as
\begin{equation}
 f_{nn}(k) = \frac{1}{k\cot \delta_S - ik} = - 2\pi^2 m_n D_{nn} (-ik) \, ,   
\end{equation}
where $k$ is the c.m. momentum.

The dressed $\Psi$ propagator packs the two-body $P_{3/2}$ interaction of $n\alpha$: 
\begin{eqnarray}
D_{n\alpha}(\kappa) =  -\frac{5}{16\pi^2 m_n}\frac{\kappa + \gamma}{\kappa^2 + \eta \kappa + \eta \gamma}   \, ,
\end{eqnarray}
where $\kappa \equiv \frac{2}{5}\sqrt{-10 m_n p_0 + \vec{p}\,^2 -i0}$. The $P_{3/2}$ partial-wave amplitude of $n\alpha$ scattering is expressed in terms of $D_{n\alpha}$:
\begin{equation}
 f_{n\alpha}(k) = \frac{1}{k\cot \delta_P - ik} = -\frac{16}{5} \frac{\pi^2 m_n \eta}{\gamma - ik}D_{n\alpha} (-ik) \, .  
\end{equation}

With these ingredients, the coupled-channel integral equations for $F_\alpha(q)$ and $F_n(q)$ are established as follows:
\begin{eqnarray}
\label{eq:Fc-faddeev}
F_{\alpha}(q) &=& 8\pi \int^{\Lambda_3}_0 q'^2 dq'\, X_{n\alpha}(q',q;B_3)\,  D_{n\alpha}(\kappa_1)\, F_{n}(q') \, ,
\\
\label{eq:Fn-faddeev}
F_{n}(q) &=& 4\pi \int^{\Lambda_3}_0 q'^2 dq'\, X_{n\alpha}(q,q';B_3)\, D_{nn}(\kappa_0)\, F_{\alpha}(q')
\nonumber\\
&&+ 4\pi \int^{\Lambda_3}_0 q'^2 dq'\, \left[X_{nn}(q,q';B_3) + \frac{q q'}{\Lambda_3^4}{H_0}(\Lambda_3)\right]\, D_{n\alpha}(\kappa_1)\, F_n(q')~,
\end{eqnarray}
where the $nn\alpha$ coupling $H_0(\Lambda_3) \equiv h \Lambda_3^4$ and the c.m. momenta for the dimeron propagators are related to $B_3$ and $q'$ as
\begin{eqnarray}
\kappa_0 &\equiv& \sqrt{m_n B_3 + \frac{3}{8}{q'}^2} \, , \\
\kappa_1 &\equiv& \frac{2}{5}\sqrt{10 m_n B_3 + 6{q'}^2} \, .
\end{eqnarray}
For simplicity, we regularize the Jacobi momentum $q'$ by an ultraviolet sharp cutoff $\Lambda_3$, while taking the cutoff values in the two-body interactions to infinities.  

\begin{figure}
\centerline{\includegraphics[scale=0.45, angle=0, clip=true]{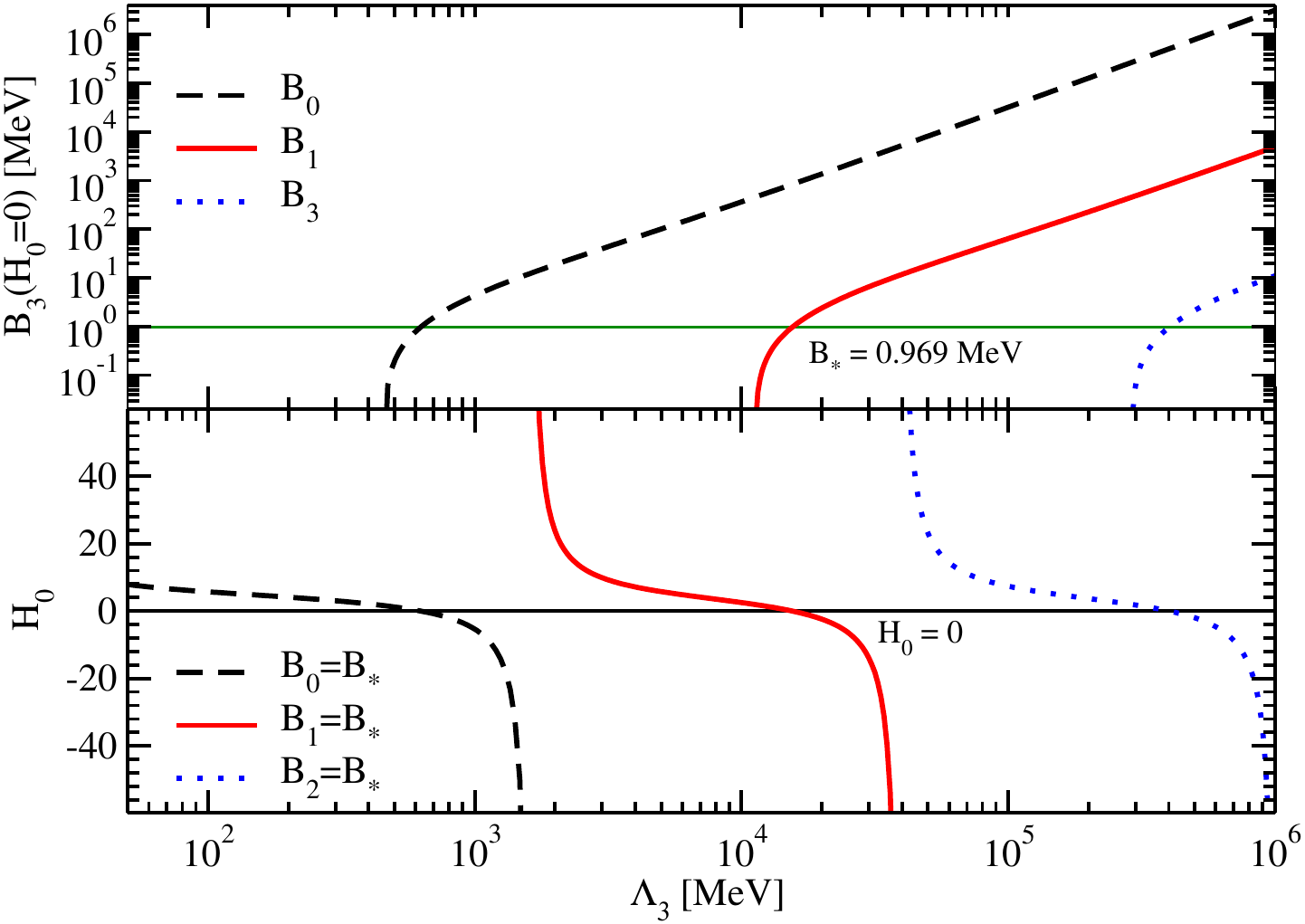}} 
\caption{(Color online) Upper panel: The three-body binding energies as functions of the cutoff $\Lambda_3$ with $H_0 = 0$. The green solid line indicates the physical binding energy $B_*$. Lower panel: ${H_0}$ as a function of the cutoff $\Lambda_3$. 
The black solid line indicates the zero axis for $H_0=0$.}
\label{pic:b3h3}
\end{figure}

We begin by solving the three-body equations without the three-body force. More specifically, we calculate the binding energies of the $nn\alpha$ system $B_3$ as a function of $\Lambda_3$ while setting the three-body force parameter $H_0(\Lambda_3) = 0$. In the upper panel in Fig.~\ref{pic:b3h3}, $B_3$ indicates a power-law dependence on $\Lambda_3$ when $\Lambda_3$ is much larger than $1/|a_0|$, $|k_{\pm}|$, and $\sqrt{m_n B_3}$. Meanwhile, a series of new bound states emerge as $\Lambda_3$ increases, and their binding energies form approximately a geometric series. These behaviors suggest that the Thomas collapse can also exist in three-body systems involving $P$-wave pairwise interactions.

Therefore, it is essential to have the $nn \alpha$ three-body interaction at LO, in order to eliminate the cutoff dependence, in the meantime preventing the Thomas collapse. By tuning $H_0$ for any given values of $\Lambda_3$, the three-body binding energy is fitted to the experimental value of $^6$He binding energy $B_* = 0.969$ MeV. In the lower panel in Fig.~\ref{pic:b3h3}, $H_0$ displays logarithmic oscillatory dependence on $\Lambda_3$, similar to the system dominated by two-body $S$-wave interactions~\cite{Bedaque:1998kg}. However, the logarithmic period of ${H}_0(\Lambda_3)$ is no longer a constant. This is because the symmetry of discrete scale invariance, which is observed in $S$-wave three-body systems, is broken by the presence of $\gamma$, an important element in the LO $n\alpha$ potential to form the $P$-wave resonance.

For lowest cutoff values where there is only one bound state, fitting to $B_*$ is straightforward. As $\Lambda_3$ increases, $H_0$ decreases to negative values and eventually diverges at $\Lambda_3 \approx 1.6$ GeV for the first time, where we begin instead to fit the first excited three-body state $B_1$ to $B_*$, 
resulting in emergence of a new $H_0$ branch from positive infinity. That is, the ground state is left to become an unphysical spurious state in the EFT. When $B_1$ is fixed, the ground-sate energy $B_0$ stays above 800 MeV, which is far beyond the EFT breakdown scale. Similar discontinuity of $H_0(\Lambda_3)$ is also shown at $\Lambda_3\approx 40$ GeV, when $B_1$ becomes too large to reproduce $B_*$, and the second excited state is used to reproduce the physical $^6$He state.

\begin{figure}
\centerline{\includegraphics[scale=0.45,angle=0,clip=true]{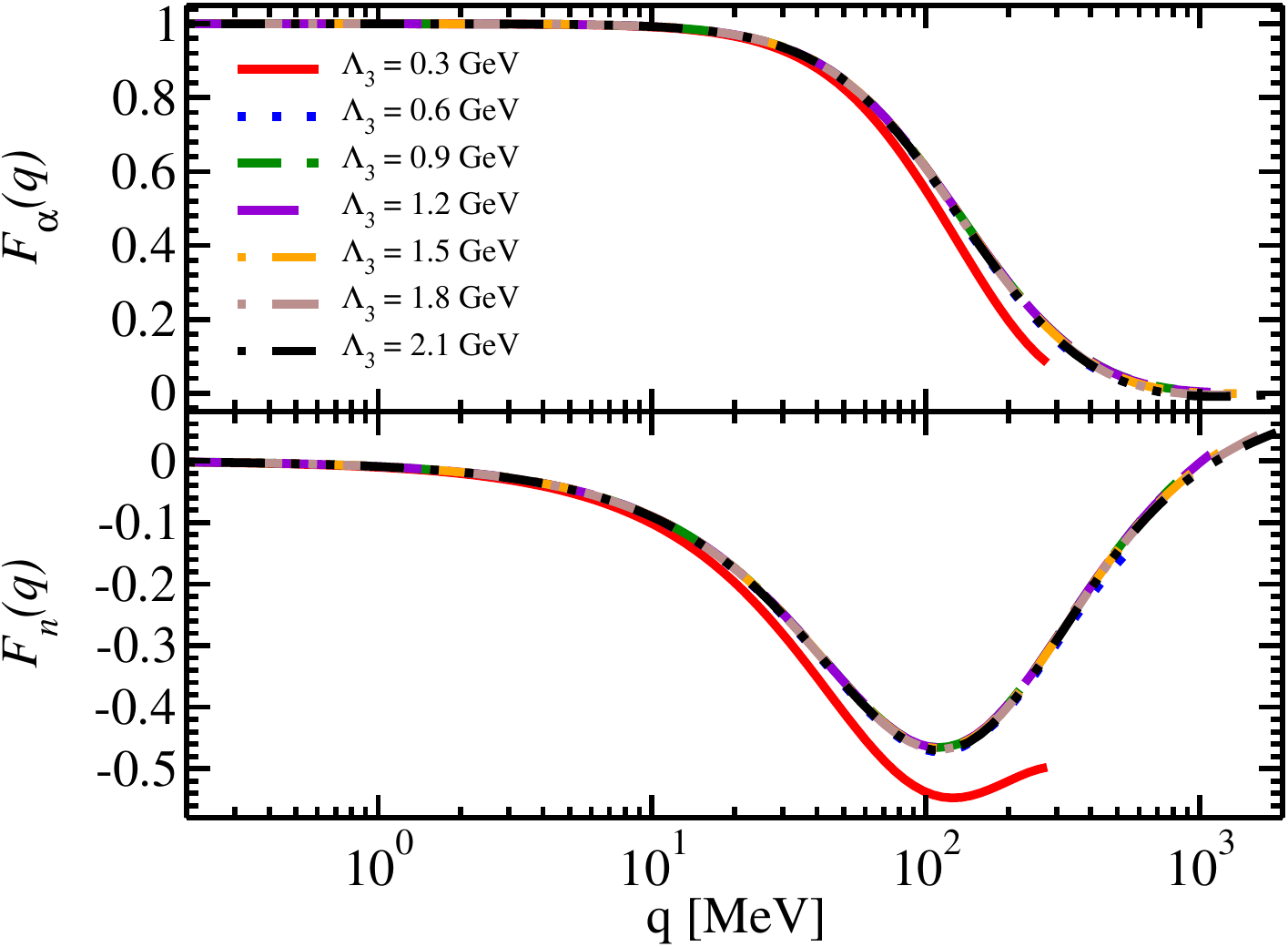}}
\caption{(Color online) The Faddeev components $F_\alpha$ and $F_n$ as functions of $q$, calculated with $\Lambda_3 = $ 0.3 GeV (red solid line), 0.6 GeV (blue dotted line), 0.9 GeV (green dash-dotted line), 1.2 GeV (purple long-dashed line), 1.5 GeV (orange dot-dashed line), 1.8 GeV (brown dot-dashed line), and 2.1 GeV (black dot-dashed line). The Faddeev components are normalized to $F_\alpha(0)=1$.}
\label{pic:FcFn}
\end{figure}

After the three-body energy is renormalized, the Faddeev components $F_n(q)$ and $F_\alpha(q)$ are calculated and checked for their cutoff dependence. In Fig.~\ref{pic:FcFn}, the Faddeev components $F_n$ and $F_\alpha$ are plotted as functions of the Jacobi momentum $q$ for a set of cutoff values. The low-momentum part of both Faddeev components exhibits an insignificant cutoff variation for $\Lambda_3 \geqslant 300 $ MeV, which suggests that the structure properties of $^6$He will be cutoff invariant at LO.

We have shown that the nonlocal, momentum-dependent $n\alpha$ interaction~\eqref{eqn:pwaveLO} can produce the same qualitative features of the $nn\alpha$ three-body system as those in Refs.~\cite{Rotureau:2012yu, Ji:2014wta} using an energy-dependent $n\alpha$ potential. The most important one is that the $nn\alpha$ three-body force must appear at LO. By replacing the energy-dependent $n \alpha$ potential with the momentum-dependent one, we can retain the fully unitary form of the $n\alpha$ dimeron propagator $D_{n\alpha}$, and in the meantime, without worrying about the unphysical redundant pole. 

\section{Summary and outlook\label{sec:summary}}

Inspired by previous works for $S$-wave interactions, we have used a momentum-dependent nonlocal potential~\eqref{eqn:pwaveLO} to develop an EFT for shallow $P$-wave resonances. The goal is to construct a framework more suitable, at least for some many-body methods, to apply in many-body calculations than previous implementations based on energy-dependent potentials.

We have shown that the potential generates the effective-range expansion with the scattering volume and generalized effective range term as the LO. Although the LO $S$ matrix has a third pole besides the resonance poles, it does not correspond to a negative-norm bound state unlike in the case of energy-dependent potentials. Higher-order ERE terms can be attained by adding systematically higher-order potentials in perturbation theory. Despite their ultraviolet (UV) suppression the potentials still need regularization, and renormalization was explained up to {\NNLO} for the two-body sector. Subleading potentials resemble those $S$-wave potentials proposed in Ref.~\cite{Beane:2021dab}, although the UV/IR symmetry of the $S$-matrix seems to be lost. 

We then applied the framework to the neutron-$\alpha$ system where the $P_{3/2}$ resonance is a prominent feature. The same scaling for the ERE parameters as in Ref.~\cite{Bertulani:2002sz} was assumed, but our framework does not generate the deep bound state as that of Ref.~\cite{Bertulani:2002sz} would. In addition, the correspondence between the Lagrangian parameters is, not surprisingly, obscured by distinctive forms of the interactions. The NLO phase shifts showed excellent agreement with the phase shift analysis from Ref.~\cite{Arndt:1973ssf}.

The neutron-neutron-$\alpha$ system was studied, with the intention to examine whether the three-body physics learned in Ref.~\cite{Ji:2014wta} can be recovered with the nonlocal potential. The answer is affirmative. The three-body force is confirmed to be at LO on renormalization ground. With the three-body parameter $H_0$ fitted to the ${}^6$He binding energy, the running of $H_0(\Lambda)$ also shows a limit-cycle behavior. However, the advantage of the momentum-dependent $n$-$\alpha$ interaction will only be fully realized in still higher-body calculations where it becomes increasingly difficult to implement energy-dependent potentials. This is under investigation.

\acknowledgments

This work was supported by the National Natural Science Foundation of China (NSFC) under Grant Nos. 11805078, 12175083 (CJ), 12275185 (BWL).

\bibliography{modhalocluster.bib}
\end{CJK*}

\end{document}